\documentclass{PoS}
\usepackage{multirow}

\title{Precise top mass determination using lepton distribution at LHC}

\ShortTitle{Precise $m_t$ determination using lepton distribution at LHC}

\author{Sayaka Kawabata%
       \\
       Institute of Convergence Fundamental Studies, \\Seoul National University of Science and Technology, Seoul 01811, Korea\\
       E-mail: \email{skawabata@seoultech.ac.kr}}

\abstract{
We present a new method to measure a theoretically well-defined mass of the top quark at the LHC.
This method uses lepton energy distribution and has a boost-invariant nature.
We perform a simulation analysis of the top mass reconstruction with this method for $t\overline{t} \to$ lepton+jets channel at the leading order.
We estimate several major uncertainties in the top mass determination and find that they are under good control.
}

\FullConference{8th International Workshop on Top Quark Physics, TOP2015\\
		 14-18 September, 2015\\
		 Ischia, Italy}

\begin{document}

\section{Introduction}
The top quark mass appears in a variety of discussions, as an important input parameter to perturbative predictions within and beyond the Standard Model (SM).
Well-known examples are electroweak precision fits~\cite{Ciuchini:2013pca} and evaluation of the SM vacuum stability~\cite{Degrassi:2012ry}.
To reduce errors in these predictions, an accurate value of the top quark mass is desired.

The top quark mass has been measured at the Tevatron and LHC using kinematic distributions of $t\overline{t}$ decay products.
Their recent combined result is $173.34 \pm 0.76$\,GeV~\cite{ATLAS:2014wva}.
However, it has been a problem that the measured mass is not well-defined in perturbative QCD.
The reason is that theoretical descriptions for the kinematic distributions of jets, which are implemented in Monte-Carlo (MC) simulations, rely on approximations and phenomenological models with tuned parameters.
In this sense, the measured mass is often referred to as ``MC mass'' and should be distinguished from theoretically well-defined masses such as the pole mass and the $\overline{\rm MS}$ mass.

The pole mass is defined as the pole position of the top quark propagator in perturbation theory.
Reflecting the fact that there is no pole in the propagator in non-perturbative QCD, the quark pole mass is sensitive to infrared physics and exhibits bad convergence properties in perturbative expansions.
It is known that this difficulty can be avoided by using short-distance masses represented by the $\overline{\rm MS}$ mass.
The relation between the pole mass and the $\overline{\rm MS}$ mass is known up to four-loop order in QCD~\cite{Marquard:2015qpa}.
The pole and $\overline{\rm MS}$ masses have been determined at the Tevatron and LHC mainly from $t\overline{t}$ cross section measurements.
The present accuracy of the top quark pole mass is around $2$\,GeV~\cite{Aad:2014kva}, which is even larger than that of the above MC mass measurements.

This situation demands a more precise determination of a theoretically well-defined top quark mass.
With the aim of determining the top quark pole mass and the $\overline{\rm MS}$ mass accurately at the LHC, we present a new method to measure the top mass.
This method uses lepton energy distribution and is called ``weight function method.''
We perform a simulation analysis of the top mass reconstruction with this method at the leading order.
This analysis will be a basis for further investigations with higher-order QCD and other corrections included.

\section{Weight function method}
The weight function method was first proposed in Ref.~\cite{Kawabata:2011gz} as a new method for reconstructing a parent particle mass, and applied to top mass reconstruction at the LHC in Ref.~\cite{Kawabataa:2014osa}.
This method has two features: using only lepton energy distribution and having a boost-invariant nature.
The former allows extraction of a theoretically well-defined top quark mass.
The latter reduces uncertainties related to top quark velocity distribution.

Using the top decay process $t \rightarrow b\ell\nu$, the top quark mass can be reconstructed only from lepton energy distribution $D(E_\ell)$.
The procedure for the top mass reconstruction with the method is as follows.
\begin{enumerate}
	\item Compute a weight function given by
	\begin{equation}
	W(E_\ell,m)=\!\int \!dE \left.\mathcal{D}_0(E;m)\frac{1}{EE_\ell} \,({\rm odd~fn.~of~}\rho)\right|_{e^{\rho}=E_\ell/E}\,,
	\label{eq:WF}
	\end{equation}
	where $\mathcal{D}_0(E;m)$ is the normalized lepton energy distribution in the rest frame of the top quark with the mass $m$.
	One can choose an arbitrary odd function for $({\rm odd~fn.~of~}\rho)$ in Eq.~(\ref{eq:WF}).
	
	\item Construct a weighted integral $I(m)$, using lepton energy distribution $D(E_\ell)$ in a laboratory frame:
	\begin{equation}
	I(m) \equiv \int dE_\ell \,D(E_\ell) \,W(E_\ell,m)\,.
	\label{eq:Im}
	\end{equation}
	
	\item Obtain the zero of $I(m)$ as the reconstructed mass:
	\begin{equation}
	I(m=m^{\rm rec}) = 0\,.
	\label{eq:zero}
	\end{equation}
\end{enumerate}
Note that the function $W(E_\ell,m)$ can be computed theoretically and the definition of the reconstructed mass corresponds to that of $m$ which you used in the calculation of the $\mathcal{D}_0(E;m)$ in step 1.
The method is based on two assumptions that the top quarks are longitudinally unpolarized\footnote{The longitudinal polarization of the top quarks produced in $t\overline{t}$ pair at the LHC is at sub-percent level~\cite{Bernreuther:2006vg}.}
and on-shell.
Deviations from these assumptions should be incorporated as corrections.

\section{Simulation analysis}

\begin{figure}[t]
	\begin{minipage}{0.45\textwidth}
		\begin{center}
			\includegraphics[width=\textwidth]{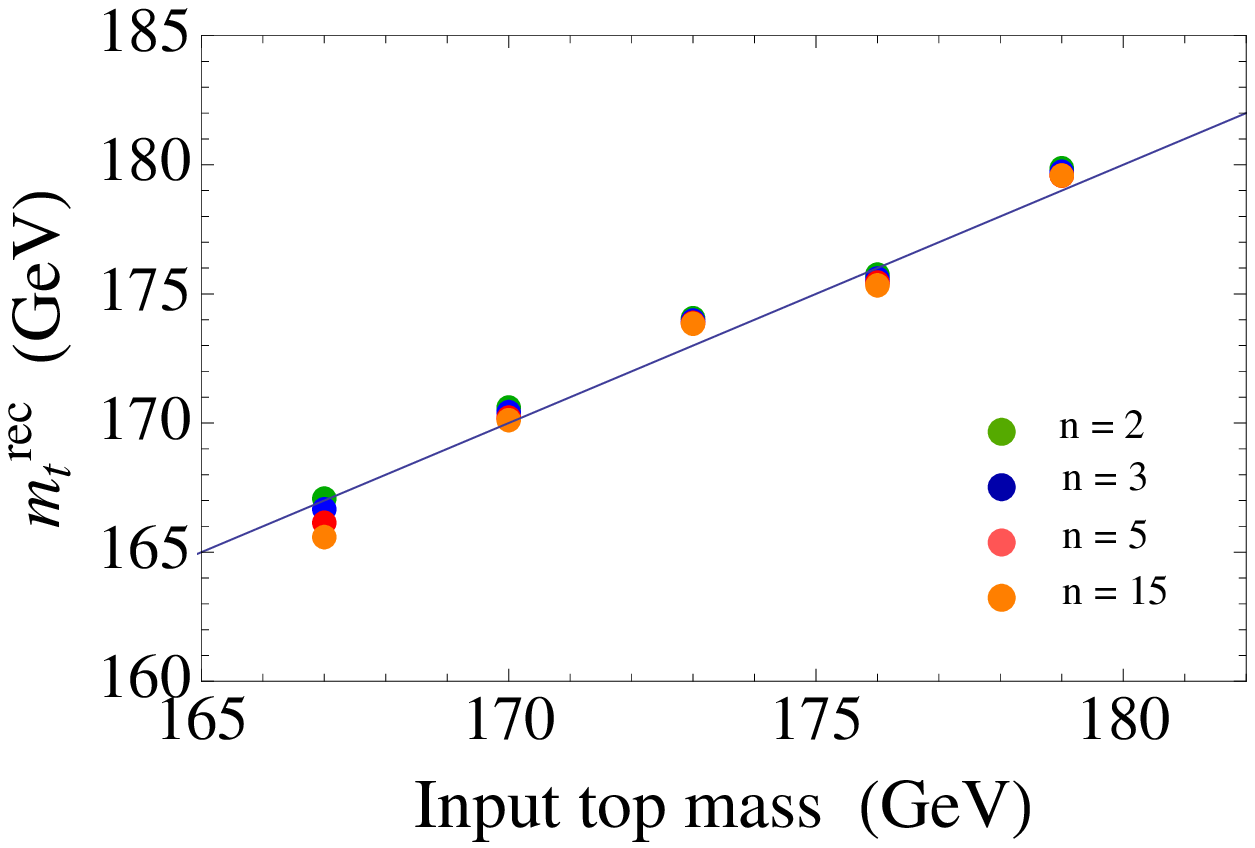}
			\caption{Reconstructed top quark mass as a function of the input mass. Weight functions used in this analysis correspond to $n=2,\,3,\,5$ and $15$ in Eq.~(3.1). The blue line shows $m_t^{\rm rec}=m_t^{\rm input}$.\label{fig:Mtrec2}}
			\label{fig:Mtrec2}
		\end{center}
	\end{minipage}
\hspace{1cm}	
	\begin{minipage}{0.45\textwidth}
	
		\begin{center}
\makeatletter
\def\@captype{table}
\makeatother		
		\vspace{-.4cm}
			\begin{tabular}{l|rc}
				\hline 
				Signal stat. error & ~~~~$0.4$&\\ \hline
				\multirow{2}{*}{Factorization scale (signal)} & $+1.5$&\\
				& $-1.4$&\\ \hline
				PDF (signal) & $0.6$ &\\ \hline
				\multirow{2}{*}{Jet energy scale (signal)} & $+0.2$ &\\ 
				& $-0.0$ &\\ \hline
				Background stat. error & $0.4$ &\\
				\hline
			\end{tabular}
			\caption{Estimates of uncertainties in GeV from several sources in the top mass reconstruction. Results for the weight function with $n=2$ are shown. The signal statistical error is for $100$\,fb$^{-1}$ and for the sum of the lepton($e,\mu$)+jets events. The background statistical error is also for $100$\,fb$^{-1}$.}
			\label{tab:uncertainties}
		\end{center}
	\end{minipage}
\end{figure}

In order to estimate experimental viability of the top mass reconstruction with the weight function method at the LHC, we perform a simulation analysis at the leading order, taking into account effects of detector acceptance, event selection cuts and background contributions.
We generate $t\overline{t} \to \mu+$jets events for the signal process.
For the background events, we consider other $t\overline{t}$, $W+$jets, $Wb\overline{b}+$jets and single-top production processes.
The center of mass energy of the LHC is assumed to be $\sqrt{s}=14$\,TeV.
We choose for the odd function of $\rho$ in Eq.~(\ref{eq:WF})
\begin{equation}
	({\rm odd~fn.~of~}\rho)\,=\,n\tanh (n\rho)/\!\cosh(n\rho)\,,
	\label{eq:oddfunc}
\end{equation}
with $n=2,\,3,\,5$ and $15$.
See Ref.~\cite{Kawabataa:2014osa} for further details of the setup of this analysis.

The lepton energy distribution $D(E_\ell)$ is deformed by various experimental effects.
In Ref.~\cite{Kawabataa:2014osa} we examined various possible sources of such experimental effects and presented a way to cope with them.
In particular, lepton cuts cause severe effects and we devised a method to solve this problem.
Fig.~\ref{fig:Mtrec2} shows the reconstructed mass as a function of the input top mass to the simulation events.
The reconstructed masses are consistent with the input masses taking into account MC statistical errors and effects of the top width in this analysis.
In Table~\ref{tab:uncertainties} we show estimates of uncertainties from several sources in the top mass determination.
The dominant source in this estimates is the uncertainty associated with factorization scale dependence.
Although the method originally had a boost-invariant nature, due to experimental effects this nature is partly spoiled.
This results in the scale uncertainty as well as the PDF uncertainty.
These uncertainties are expected to be reduced by including next-to-leading order corrections in this method.

\section*{Acknowledgements}
I would like to thank Y. Shimizu, Y. Sumino and H. Yokoya for fruitful collaborations.
This research was supported by Basic Science Research Program through the National Research Foundation of Korea (NRF) funded by the Ministry of Science, ICT and Future Planning (Grant No. NRF-2014R1A2A1A11052687).


\begin{thebibliography}{99}

\bibitem{Ciuchini:2013pca} 
  M.~Ciuchini, E.~Franco, S.~Mishima and L.~Silvestrini,
  JHEP {\bf 1308}, 106 (2013)
  [arXiv:1306.4644 [hep-ph]];
%
  M.~Baak {\it et al.} [Gfitter Group Collaboration],
  Eur.\ Phys.\ J.\ C {\bf 74}, 3046 (2014)
  [arXiv:1407.3792 [hep-ph]];
%
  G.~F.~Giudice, P.~Paradisi and A.~Strumia,
  JHEP {\bf 1511}, 192 (2015)
  [arXiv:1508.05332 [hep-ph]].

\bibitem{Degrassi:2012ry} 
  G.~Degrassi, S.~Di Vita, J.~Elias-Miro, J.~R.~Espinosa, G.~F.~Giudice, G.~Isidori and A.~Strumia,
  JHEP {\bf 1208}, 098 (2012)
  [arXiv:1205.6497 [hep-ph]];
%
  D.~Buttazzo, G.~Degrassi, P.~P.~Giardino, G.~F.~Giudice, F.~Sala, A.~Salvio and A.~Strumia,
  JHEP {\bf 1312}, 089 (2013)
  [arXiv:1307.3536 [hep-ph]].

\bibitem{ATLAS:2014wva} 
  [ATLAS and CDF and CMS and D0 Collaborations],
  arXiv:1403.4427 [hep-ex].
  
\bibitem{Marquard:2015qpa} 
  P.~Marquard, A.~V.~Smirnov, V.~A.~Smirnov and M.~Steinhauser,
  Phys.\ Rev.\ Lett.\  {\bf 114}, no. 14, 142002 (2015)
  [arXiv:1502.01030 [hep-ph]].
  
\bibitem{Aad:2014kva} 
  G.~Aad {\it et al.} [ATLAS Collaboration],
  Eur.\ Phys.\ J.\ C {\bf 74}, no. 10, 3109 (2014)
  [arXiv:1406.5375 [hep-ex]];
%
  S.~Chatrchyan {\it et al.} [CMS Collaboration],
  Phys.\ Lett.\ B {\bf 728}, 496 (2014)
  [Phys.\ Lett.\ B {\bf 728}, 526 (2014)]
  [arXiv:1307.1907 [hep-ex]];
%
  G.~Aad {\it et al.} [ATLAS Collaboration],
  JHEP {\bf 1510}, 121 (2015)
  [arXiv:1507.01769 [hep-ex]].

\bibitem{Kawabata:2011gz} 
  S.~Kawabata, Y.~Shimizu, Y.~Sumino and H.~Yokoya,
  Phys.\ Lett.\ B {\bf 710}, 658 (2012)
  [arXiv:1107.4460 [hep-ph]];
  S.~Kawabata, Y.~Shimizu, Y.~Sumino and H.~Yokoya,
  JHEP {\bf 1308}, 129 (2013)
  [arXiv:1305.6150 [hep-ph]].

\bibitem{Kawabataa:2014osa} 
  S.~Kawabata, Y.~Shimizu, Y.~Sumino and H.~Yokoya,
  Phys.\ Lett.\ B {\bf 741}, 232 (2015)
  [arXiv:1405.2395 [hep-ph]].

\bibitem{Bernreuther:2006vg} 
  W.~Bernreuther, M.~Fuecker and Z.~G.~Si,
  Phys.\ Rev.\ D {\bf 74}, 113005 (2006)
  [hep-ph/0610334];
%
  W.~Bernreuther and Z.~G.~Si,
  Phys.\ Lett.\ B {\bf 725}, 115 (2013)
  [Phys.\ Lett.\ B {\bf 744}, 413 (2015)]
  [arXiv:1305.2066 [hep-ph]].

  
\end{thebibliography}
\end{document}